\documentclass{aa}
\usepackage{epsfig,amsmath,amssymb}

\def\msol{M_\odot}
\def\msun{M_\odot}
\def\mjup{M_{\rm J}}
\def\rjup{R_{\rm J}}
\def\mearth{\,{\rm M}_\oplus} 

\def\mcore{\,{\rm M}_{\rm core}}
\def\zenv{\,{\rm Z}_{\rm env}}

\def\simgr{\,\hbox{\hbox{$ > $}\kern -0.8em \lower 1.0ex\hbox{$\sim$}}\,}
\def\simle{\,\hbox{\hbox{$ < $}\kern -0.8em \lower 1.0ex\hbox{$\sim$}}\,}
\def\beq{\begin{equation}}
\def\eeq{\end{equation}}

\begin{document}

\title{Birth and fate of hot-Neptune planets}

 \author{I. Baraffe\inst{1,2}, Y. Alibert \inst{3}, G. Chabrier\inst{1,2}, W. Benz\inst{3}
}

\offprints{I. Baraffe}

\institute{C.R.A.L,
 Ecole Normale Sup\'erieure, 46 all\'ee d'Italie, 69007 Lyon, France (ibaraffe, chabrier@ens-lyon.fr)
\and
International Space Science Institute, Hallerstr. 6, CH-3012, Bern, Switzerland
\and
Physikalisches Institut, University of Bern, Sidlerstr. 5, CH-3012,Bern, Switzerland
(yann.alibert, willy.benz@phim.unibe.ch)
}

\date{Received /Accepted}

\titlerunning{Birth and fate of hot-Neptune planets: a consistent formation-evolution scenario}
\authorrunning{Baraffe et al.}
\abstract{This paper presents a consistent description of the formation and the subsequent evolution of gaseous planets, with special attention to short-period, low-mass hot-Neptune planets characteristic of  $\mu$ Ara-like systems. We show that core accretion including migration and disk evolution and subsequent evolution taking into account irradiation and evaporation provide a viable formation mechanism for this type of strongly irradiated light planets. At an orbital distance $a \, \simeq$ 0.1 AU,
this revised core accretion model leads to the formation of planets with total masses ranging from $\sim$ 14 $\mearth$ (0.044 $\mjup$) to $\sim$ 400 $\mearth$ (1.25 $\mjup$). The newly born planets have a dense core of $\sim$ 6 $\mearth$, independent of the total mass, and heavy element enrichments in the envelope,  $M_{\rm Z,env}/M_{\rm env} $, varying from 10\% to 80\% from the largest to the smallest planets. We examine the dependence of the evolution of the born planet on the evaporation rate due to the incident XUV stellar flux. In order to reach a $\mu$ Ara-like mass ($\sim$ 14 $\mearth$) after $\sim $ 1 Gyr, the initial planet mass must range from 166 $\mearth$ ($\sim$ 0.52 $\mjup$) to about 20 $\mearth$, for evaporation rates varying by 2 orders of magnitude, corresponding to 90\% to 20\% mass loss during evolution. 
The presence of a core and heavy elements in the envelope affects appreciably the structure and the evolution of the planet and yields $\sim 8\%-9\%$ difference in radius compared to coreless objects of solar composition for Saturn-mass planets.
 These combinations of evaporation rates and internal compositions
 translate into different detection probabilities, and thus different statistical distributions for hot-Neptunes and hot-Jupiters.
These calculations provide an observable diagnostic, namely a mass-radius-age relationship to distinguish between the present core-accretion-evaporation model
and the alternative colliding core scenario for the formation of hot-Neptunes.

}

\maketitle

\section{Introduction}

Since the discovery of the first exoplanet by Mayor \& Queloz (1995), planet hunters have been discovering planets with smaller and smaller masses. With the discovery of objects in the range of 10-20 $\mearth$, i.e. 0.03-0.06 $\mjup$ \footnote{1 $\mjup = 318 \mearth$.} (Butler et al. 2004; McArthur et al. 2004; Santos et al. 2004), a new step in the quest for Earth-like planets has been taken. While representing an extraordinary achievement from the observational side, theorists still struggle to understand the structure and the origin of these light giant planets. Are they essentially composed of ices and rocks,
with possibly a thin atmosphere, like our ice giants Uranus and Neptune? Or do they originate from larger gaseous planets, with a large gaseous envelope and a relatively small central rocky core? The answer to these questions requires an understanding of their formation process.

Current planet formation scenarios, based either on the core accretion model or on gravitational instability, can more or less explain the presence of relatively massive planets with masses $\simgr \, 100  \, \mearth$ (or even larger in the case of the gravitational instability scenario) at various orbital separations but they do not necessarily predict the formation of a large number of lighter planets (Boss 2001; Ida \& Lin 2005; Papaloizou \& Nelson 2005).

In the framework of the core accretion model (Pollack et al. 1996), the general expectation was to find preferentially planets either less massive or more massive than the newly discovered Neptune-like planets, for these latter lie within the domain of critical mass ($\sim 10-20 \mearth$) above which runaway accretion of gas begins. Similarly, Ida \& Lin (2004) suggest a possible deficit of intermediate mass planets ($\sim 10-100 \mearth$, i.e. $\sim 0.03-0.3 \mjup$) with orbital separation $a<3$ AU. On the opposite, using N-body simulations of colliding cores in a protoplanetary disk, Brunini \& Cionco (2005) find that Neptune-like planets close to their host-star can form easily as a by-product of planetary formation. If their scenario is correct, these planets should be of ice-rock composition with only a thin atmosphere. They also predict that a large population of these "hot cores" should be discovered in a near future.

As an alternative to this colliding core scenario, and given the close orbital distance of the Neptune-like planets  discovered up to now, one cannot  exclude the possibility that they formed initially as larger giant planets wich have undergone atmospheric evaporation during their lifetime (Baraffe et al. 2005). Within this picture, formation and evolution are strongly correlated: a correct understanding of light planet properties thus requires a consistent description of the planet formation and evolution in order to interpret present-day observations.

This paper is a first attempt to derive such a consistent picture from the planet formation to its subsequent evolution. We apply the core accretion models (cf. \S 2) developed recently by Alibert et al. (2005a) to the 14 $\mearth$ (0.044 $\mjup$) planet (modulo sin $i$) orbiting around the G-star $\mu$ Ara at an orbital distance $a=0.09$ AU (Santos et al. 2004). We focus on $\mu$ Ara-like planets  because it is still reasonable to apply the Alibert et al. (2005a) formation model at this orbital distance, whereas it is no longer the case at the location of the two other Neptune-like planets
which are located at $a \, < \, 0.05$ AU from their parent star (Butler et al. 2004; McArthur et al. 2004). Indeed, for $a \simle 0.1$ AU, the description of the inner part of the disk, including tidal and magnetic interactions with the star, is too crude to provide a reliable formation scenario.

Adopting the main characteristics of the newly born planets as predicted by the Alibert et al. (2005a) formation model (core mass, heavy element content), we follow the later evolution  of these planets, according to Baraffe et al. (2004, 2005), taking into account irradiation and evaporation effects due to the vicinity of the parent star. We examine the sensitivity of the results on the evaporation rate by exploring a range of different rates and present our results in \S 4. Predictions and uncertainties of our scenario are discussed in \S 5.

\section{Formation model}

The adopted formation model is an extension of the core accretion model for giant gaseous planets developed by Pollack et al. (1996). The present model includes the effect of migration and of  disc evolution on the planet formation. Details can be found in Alibert et al. (2005a). The model describes successfully the formation of Jupiter-like planets on timescales consistent with typical disk lifetimes (Alibert et al. 2004). When applied to the formation of Jupiter and Saturn, the model yields core masses and heavy element contents in the envelope in agreement with recent interior structure models for Jupiter and Saturn (Saumon \& Guillot 2004).
Moreover, the model predicts surface compositions in agreement with both {\it in situ} abundance measurements by the {\it Galileo} probe for Jupiter, and with Earth based measurements for Saturn (Alibert et al. 2005b).

We employ this formation model to determine the properties of an ensemble of planets which can form eventually at 0.1 AU of a solar type star, like the planet found around $\mu$ Ara.
We performed a large number of simulations, similar to those described in Alibert et al. (2005a, see their \S 3.2), exploring different initial parameters (orbital separation, dust-to-gas ratio in the disc, photoevaporation rate, disc mass). 
The initial disk profile is assumed to be a power law with a gas surface density
$\Sigma = \Sigma_0 (r / 5.2 AU) ^{-3/2}$, with
$\Sigma_0 = 100, 300, 500, 700$ g/cm$^2$. The surface
density of planetesimals is $\Sigma_{\rm P} = f_{D/G} \Sigma$,
where the dust-to-gas ratio $f_{D/G}$ is equal to $1/30$ for temperatures
below 150 K, and $1/120$ otherwise. These values reflect the
high metallicity of the central star in the $\mu$ Ara system.
Different initial orbital separations of the planet are adopted: $a_{\rm ini} = 0.2,
0.3, 0.4, 0.6, 1, 1.3, 1.6, 1.9, 2.2, 2.5, 3, 3.5, 4, 4.5, 5, 6, 7$ AU.
The evolution of the disk gas surface density is calculated
according to the $\alpha$ prescription (Shakura \& Sunyaev 1973),
with $\alpha = 2 \cdot 10^{-3}$, including a photoevaporating
term (see Alibert et al. 2005a). Different values for the total photoevaporation rate
are assumed:  $10^{-9} \msun/$yr, $2 \times 10^{-9} \msun/$yr, $4 \times
10^{-9} \msun/$yr,
$8 \times 10^{-9} \msun/$yr and $15 \times 10^{-9} \msun/$yr. Finally,
we assume a time delay between the beginning of disk evolution
and the planet formation process. We adopt different time delays: 
0 Myr, 0.5 Myr, 1 Myr, 1.5 Myr, 2 Myr, 3 Myr, 4 Myr, 6 Myr and 8 Myr.
The simulations include migration with a  migration rate for low mass planets (type I
migration, see Ward 1997) based on the
results of Tanaka et al. (2002), reduced by a
constant factor equal to 0.01 (see Alibert et al. 2005a). As shown in Alibert et al. (2005b) for the cases of Jupiter and Saturn, a variation of the type I migration rate does not affect the final structure of the planet.
Each simulation produces a planet, eventually falling
onto the central star, and we focus on  the properties of planets ending their formation around 0.1 AU, whatever their final mass. 
The results are summarised  in Figure 1 which displays the core mass $M_{\rm core }$ and the mass fraction of heavy elements in the envelope $Z_{\rm env}=M_{\rm Z,env}/M_{\rm env} $ as a function of the total mass of the planet.
$M_{\rm Z,env}$ is the amount of mass deposited by the infalling planetesimals in the envelope whereas $M_{\rm core }$ is the mass of solids that directly reach the central solid core of the planet.
 The mass of the core depends on some poorly known quantities,
such as the characteristic size of the infalling planetesimals
(we assume here a size of 100 km), and their internal properties
(e.g. tensil strength, for which  we use here
values relevant for water ice). Note, however, that the {\it total content of heavy
elements} is much less sensitive to these quantities. As shown
in \S 4.2,  the effect of evaporation
depends mainly on the mean heavy element
enrichment of the planet, and not on their precise distribution in the planet.
Finally, we note that we do not take into account in the formation calculations the modifications
of opacity and equation of state due to
changes of the chemical composition in the envelope (see discussion in \S 5.2).

\begin{figure}
\psfig{file=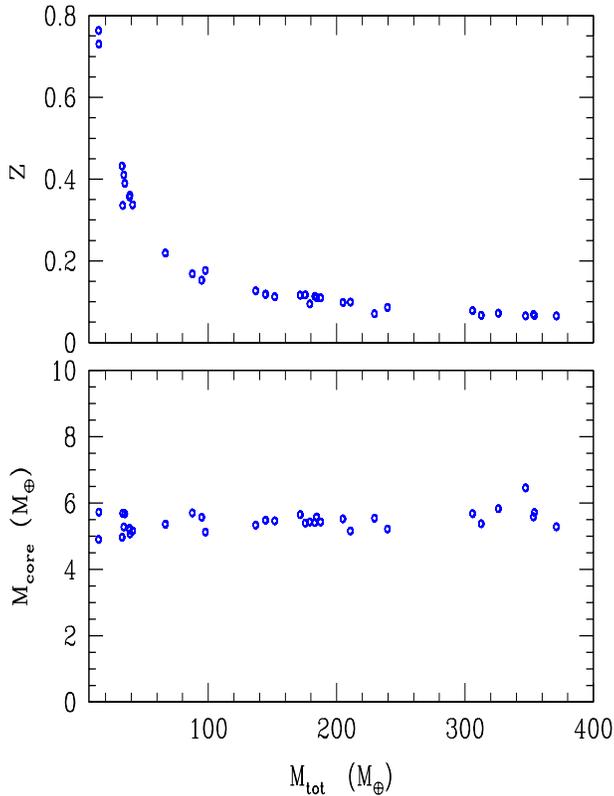,height=120mm,width=88mm} 
\caption{Properties of planets forming at 0.1 AU from their sun, based on a set of simulations using different initial conditions (see text, \S 2). {\it Lower panel:} 
Core mass as a function of total mass. {\it Upper panel:} Metallicity in the envelope 
$Z_{\rm env}$, defined as the ratio of heavy element mass in the envelope $M_{\rm Z,env}$ to the total envelope mass $M_{\rm env}$, as a function of total mass.}
\label{fig1}
\end{figure}

The simulations predict  that planets reaching a total mass in the range  14-400 $\mearth$ ($\sim 0.04-1.25 \, \mjup$) form with essentially the same core mass $M_{\rm core} \simeq 6 \mearth$ (cf. Fig. 1). The heavy element content in the envelope is found to increase substantially with decreasing total mass. The envelope metallicity mass fraction $Z_{\rm env}$ varies from 80\% for a 14 $\mearth$ planet to 10\% for masses $>$ 100 $\mearth$. Note  that our calculations {\it do not} take into account the probability of occurrence of the initial parameters leading to the formation of these  planets. 
Therefore, the number of light planets that can actually form by this extended core accretion mechanism and its fraction compared to the number of more massive planets cannot be inferred from these calculations.

\section{Evolutionary models}

\subsection{Irradiation and evaporation effects}

Using the initial conditions provided by the planet formation model described in \S 2, core mass and heavy element enrichment for a total planet mass, we calculate the subsequent evolutionary sequences for different values of these initial planetary masses. Evolutionary models take into account irradiation effects from the parent star on the planet atmospheric and internal structure, as described in Baraffe et al. (2003) and Chabrier et al. (2004). The incident stellar flux received by the planet atmosphere is determined from the orbital and stellar parameters characteristic of the $\mu$-Ara system ($a$ = 0.09 AU, $m_\star$ = 1 $\msol$, $R_\star \, \simeq \, 1 \, R_\odot$, T$_{\rm eff \star} \, \simeq \, 5800$ K, cf. Santos et al. 2004). Evaporation effects are also included, as described in Baraffe et al. (2004), using, as our fiducial model, the energy-limited escape model derived by Lammer et al. (2003, hereafter L03). This model describes a hydrodynamic mass loss process due to the high energetic XUV flux of the parent star. We  also consider smaller evaporation rates. Indeed, L03 assume that the planet undergoes maximal energy-limited evaporation, thus providing  an upper limit for escape rates. Their predicted escape rate for HD 209548b is $\sim$ 100 times larger than the {\it lower limit} measured by Vidal-Madjar et al. (2003) for this planet. In a recent study, Yelle (2004) included a better treatment of atmospheric chemistry, a crucial ingredient for handling cooling processes, and found escape rates 20 or 100 times smaller than the ones estimated by L03. Taking into account a two-dimensional, hydrodynamical energy deposition calculation instead of the single-layer heating model used in L03, Tian et al. (2005) find escape rates for HD 209458b at least $\sim$ 16 times smaller than  L03. Finally, in a recent study, Jaritz et al. (2005) analyze Roche-lobe effects on expanded upper atmospheres of close-in giant planets. They conclude that, in some cases, planets can undergo geometrical blow-off, rather than classical hydrodynamic blow-off\footnote{Classical dynamical blow-off occurs above a critical temperature for which the thermal velocity of atoms or molecules in the planetary upper atmosphere exceeds their escape velocity.
In contrast, the geometrical blow-off, as defined by Lecavelier des Etangs et al. (2004), occurs when the upper atmosphere reaches the planet Roche Lobe.},
because  the critical level where blow-off occurs cannot be reached before the exobase level reaches the Roche-Lobe. Jaritz et al. (2005) show that the transit planet OGLE-TR 26b is in such a configuration and find an escape rate 25 times lower than the maximal energy-limited mass loss calculated by L03. Given the large uncertainties in the escape rate illustrated by these different studies, we performed evolutionary calculations with 1, 1/20 and 1/100 times 
the escape rate based on the model of L03,  corresponding 
to values ranging from 
5$\times 10^{12}$ g/s (2.6 $10^{-8}$ $\mearth$/yr) to 
 7$\times 10^{9}$ g/s  (4 $10^{-11}$ $\mearth$/yr) for planets older than 1 Gyr.

\subsection{Models with a rocky core}

All our previous evolutionary calculations  were done for coreless planets (Baraffe et al. 2003, 2004, 2005, Chabrier et al. 2004). In order for the evolution to be consistent with
our formation model, we have now included a central rocky core in our structure and evolution calculations. Instead of using a constant core density, as done in some exo-planet models (e.g Bodenheimer et al. 2003), we have implemented the ANEOS equation of state (Thompson \& Lauson 1972). This equation of state (EOS) describes the thermodynamic properties of different material of planetary interest (ice, dunite, iron), derived from the Helmholtz free energy. Even though this EOS has physical limitations\footnote{e.g electrons are treated classically, i.e no degeneracy; see also the discussion in \S 3.3.4 of Guillot et al. 2004}, it gives a thermodynamic description of phase transitions and thermodynamic quantities relevant for the evolution of planets as well as Rosseland opacities and electron conductivities. In the evolutionary code, we integrate the structure equations from the center to the surface and at the core/envelope boundary, the EOS switches from  ANEOS to the Saumon-Chabrier EOS (Saumon, Chabrier \& VanHorn 1995).
At this boundary, a density jump is present due to the change in composition but continuity in pressure and temperature is enforced. In order to check our implementation of ANEOS, we have compared the mass-radius relationships obtained for water ice and olivine (or dunite, Mg$_2$SiO$_4$) planets to the results displayed in Guillot (2005) and Guillot et al. (1996) based on the same EOS. We find an excellent agreement with these studies.

As suggested by the formation model (cf. Fig. 1), we adopt in the following a core mass of 6 $\mearth$ independent of  the initial planet mass. We assume that the core is made of dunite, as representative of rock. This material yields  typical mean densities in the core
$\sim$ 6-7 g cm$^{-3}$.  We also performed comparative  calculations with water ice cores, 
corresponding to a lower mean density $\sim$ 3 g cm$^{-3}$. Adopting ice or dunite
for  the core composition slightly changes the mass-radius relationship for planets of identical core and total mass. However, this change has no effect on the main conclusion of our paper and we therefore adopt dunite cores in the following.

\subsection{Heavy element enrichment in the envelope}

As shown in Fig. 1, heavy element enrichment in the envelope due to accretion of planetesimals during the formation process can be
significant, up to 80\% in mass for the lightest planet formed (see Alibert et al. 2005a for details on the calculation of the interaction between infalling planetesimals and the envelope,
and the resulting heavy element enrichment). Under the conditions of temperature and pressure characteristic of planet interiors, the contribution to the EOS due to the interactions between particles is of the order of
or even dominate the kinetic (perfect gas) contribution. However, an accurate treatment of the EOS of such an {\it interacting} multi-species fluid is elusive and approximations are necessary.
Since our goal is to explore the effect of a heavy element contamination on the planet
structure and evolution, we treat this contribution to the EOS as follows. The presence of heavy elements with a mass fraction $Z$ in the envelope is mimicked by an equivalent helium fraction $Y_{\rm equiv}$ = $Y+Z$ in the H/He Saumon-Chabrier EOS, where  $Y$ is the genuine helium mass fraction in the envelope. This procedure has been discussed in Chabrier et al. (1992) for models of Jupiter and Saturn and in Chabrier \& Baraffe (1997), who demonstrate its validity for solar metallicity i.e. for $Z \ll 1$. It is clear that for metal enrichment as high as obtained by the formation model, such an approximation becomes extremely crude. Since the heavy element enrichment will dominantly affect the {\it structure} of the planet (the thermal contribution from the solid core is small) and since within first order the different contributions to the EOS scale with the density, this approximation, however, is good enough to give a semi-quantitative description of the effect of an increase of $Z$ on the thermodynamic properties of the envelope material, and thus on the mechanical properties (mass-radius relationship and, hence, mean density) of our planets.

 \section{Results}

 We followed the evolution of planets of different initial masses, $m_{\rm i}$, in the range $\sim$ 18-175 $\mearth$ (0.055 to 0.55 $\mjup$).  The initial interior structures $\mcore$ and $\zenv$
for the evolutionary models are the ones obtained from the formation model. We adopt an initial radius which is arbitrarily large but lies well within the Roche lobe radius of the planet. Work is in progress to better determine the final radius of the new born planet once accretion has terminated in order to use it as a more accurate initial condition for the evolutionary models. It should be stressed, however, that no planet formation model can provide an accurate determination of the final planet radius, at the end of the gas accretion phase. Indeed, the calculation of this final radius is a daunting problem, which depends crucially on the accretion process, not expected to be spherical, and  on the ability of the accreting planet to radiate away its thermal energy, which requires a correct description of the (accretion) radiative shock.
According to the results of the formation model, we assume that all planets have a rocky (dunite) core mass of 6 $\mearth$ and a heavy element enrichment corresponding to their mass (see Fig. 1). 
Evaporation rates are varied between 1 and 1/100 times the rate given by L03.  We focus on sequences which, for a fixed evaporation rate, reach a mass close to a Neptune-mass (17 $\mearth$ or 0.053 $\mjup$) within a few billion years, the inferred age for $\mu$-Ara (Santos et al. 2004; Baraffe et al. 2005). Table \ref{summary} summarizes the different evolutionary sequences presented below, with  their inputs and parameters.

\begin{table*}
\caption{Properties of evolutionary sequences with different initial masses,
escape rate prescriptions, core masses and envelope metallicities, which
 can reach a mass close to the Neptune mass 
in more than 1 Gyr.}            
\begin{tabular}{llcccccc}
\hline\noalign{\smallskip} 
Initial mass & escape rate & $\mcore$/$\mearth$ & $\zenv$ & $\log$ t (yr) & M/M$_\oplus$ (M$_J$) & R/$\rjup$ &  $\dot {\rm M}$ ($\mearth$/yr) \\
\noalign{\smallskip}
\hline
\hline
\noalign{\smallskip}
166 $\mearth$  & L03 & 0 & 0.02 
 & 7.0 & 166. (0.52) &  1.44 & 3.90 10$^{-7}$ \\ 
 &&&&  8.0 & 137. (0.43) &  1.28 & 3.32 10$^{-7}$\\
 &&&&  9.0 &  32. (0.10) &  1.25 & 1.18 10$^{-7}$\\
 &&&&  9.03 &  20. (0.06) &  1.40 & 2.45 10$^{-7}$\\
 \hline
166 $\mearth$  & L03 & 6 & 0.02
 & 7.0 & 166. (0.52) &  1.39 & 3.48 10$^{-7}$\\
 &&&& 8.0 & 140. (0.44) &  1.23 & 2.85 10$^{-7}$ \\
 &&&&  9.0 &  70. (0.22) &  1.08 & 3.57 10$^{-8}$ \\
 &&&& 9.44 &  21. (0.07) &  0.94 & 2.92 10$^{-8}$ \\
\hline
166 $\mearth$  & L03 & 6 & 0.1
 & 7.0 & 166. (0.52) &  1.33 & 3.09 10$^{-7}$\\
 &&&&  8.0 & 143. (0.45) &  1.18 & 2.47 10$^{-7}$\\
 &&&&  9.0 &  86. (0.27) &  1.05 & 2.69 10$^{-8}$\\
 &&&& 9.80 &  18. (0.06) &  0.83 & 1.07 10$^{-8}$\\
\hline
50 $\mearth$  & L03/10 & 6 & 0.2
  & 7.0 &  50. (0.15) &  1.52 & 1.5 10$^{-7}$\\
  &&&& 8.0 &  41. (0.13) &  1.19 & 8.83 10$^{-8}$\\
  &&&& 9.0 &  23. (0.07) &  0.93 & 6.84 10$^{-9}$\\
  &&&& 9.23 &  19. (0.06) &  0.87 & 3.99 $10^{-9}$\\
\hline
33 $\mearth$  & L03/20 & 6 & 0.4
 &    7.0 &  33. (0.10) &  1.31 & 7.30 10$^{-8}$\\
 &&&& 8.0 &  28. (0.09) &  1.08 & 4.79 10$^{-8}$\\
 &&&&  9.0 &  20. (0.06) &  0.82 & 2.72 10$^{-9}$\\
 &&&& 10.0 &  15. (0.05) &  0.64 & 2.02 10$^{-10}$\\
\hline
18 $\mearth$  & L03/100 & 6 & 0.4
 &    7.0 &  18. (0.06) &  1.35 & 3.01 10$^{-8}$\\
 &&&&  8.0 &  16. (0.05) &  1.03 & 1.43 10$^{-8}$ \\
 &&&&  9.0 &  14. (0.04) &  0.76 & 6.02 10$^{-10}$ \\
 &&&& 10.0 &  13. (0.04) &  0.61 & 4.02 10$^{-11}$\\
\hline \hline
  \end{tabular}
  \label{summary}
  \end{table*}

\subsection{Evolution with maximal escape rate}

\begin{figure}
\psfig{file=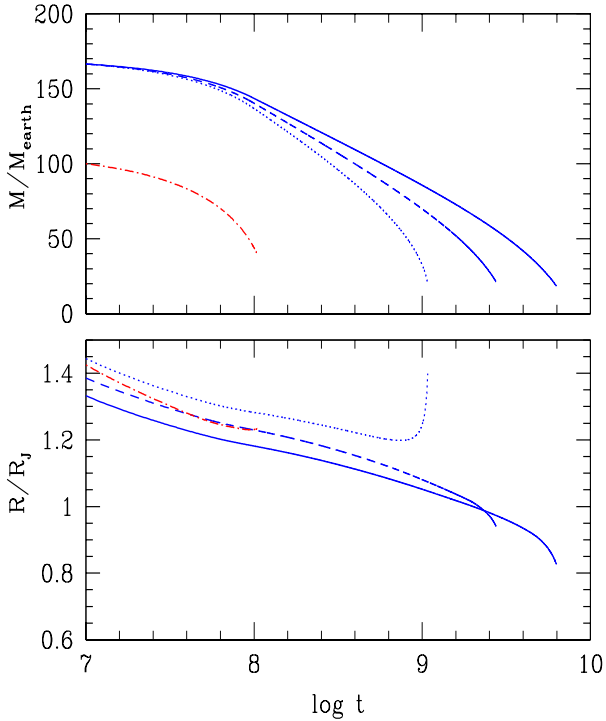,height=110mm,width=88mm}
\caption{Evolution of irradiated and evaporating planets at an orbital distance $a$=0.09 AU from a G-type star. The escape rate of L03 is used. Sequences with initial mass  $\sim$ 166 $\mearth$ (0.52 $\mjup$) are shown for
different cases:  $\mcore$=0 and metallicity in the envelope $Z_{\rm env}$=$Z_\odot$ (dot); $\mcore$=6 $\mearth$ and $Z_{\rm env}$=$Z_\odot$
(dash); $\mcore$=6 $\mearth$ and $Z_{\rm env}$=0.1 (solid). The (red) dash-dotted curve corresponds to an initial mass  $\sim$ 100 $\mearth$ (0.3 $\mjup$), with  $\mcore$=6 $\mearth$ and $Z_{\rm env}$=0.15,
as suggested by Fig. 1.}
\label{fig2}
\end{figure}

\begin{figure}
\psfig{file=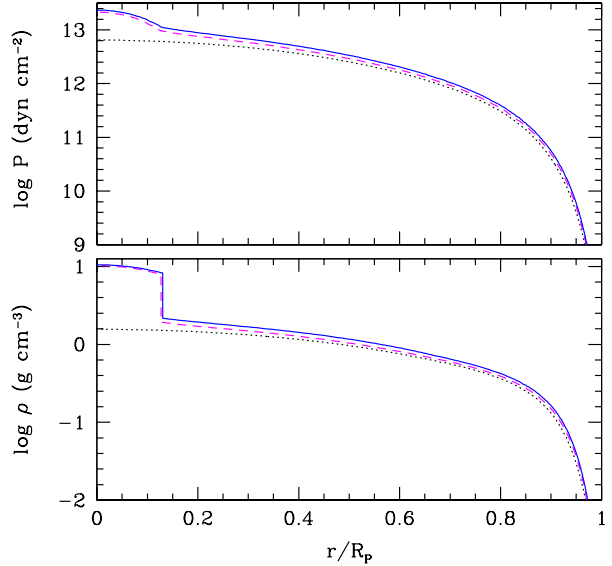,height=88mm,width=88mm}
\caption{Pressure (upper panel) and density (lower panel) radial profiles (in units of total radius) for a planet with mass  $\sim$ 166 $\mearth$ (0.52 $\mjup$). The structures displayed  correspond to  an age of 1.5 Gyr. Different cases are shown: $\mcore$=0 and metallicity in the envelope $Z_{\rm env}$=$Z_\odot$ (dot); $\mcore$=6 $\mearth$ and $Z_{\rm env}$=$Z_\odot$ (dash); $\mcore$=6 $\mearth$ and $Z_{\rm env}$=0.1 (solid).}
\label{figpro}
\end{figure}

In the case of  the maximal evaporation rate (rate from L03,) the initial masses must be greater than about 0.5 $\mjup$ (160 $\mearth$) to produce a $\mu$-Ara like Neptune-mass planet after a few billion years of evolution, and $\simgr$ 0.3 $\mjup$, about a Saturn-mass, to reach this limit after about 100 Myr. Figure \ref{fig2} shows  evolutionary sequences of planets with initial mass $m_{\rm i}$=0.3 $\mjup$ and 0.52 $\mjup$. The results for $m_{\rm i}$=0.52 $\mjup$ (166 $\mearth$) illustrate the significant effect of the presence of a core and of the envelope metal enrichment on the planet evolution. Both effects yield an increased average metallicity and thus an increase of the mean density of the planet.
In order to highlight the effect of the presence of
a central core and heavy element enrichment on the structure of a planet, Table \ref{struc} compares, for {\it constant mass} evolution, i.e. without evaporation, the radius of a 0.3 $\mjup$, i.e. Saturn-mass, coreless gaseous planet of solar composition and the one of a $\mcore=6\,\mearth$ and $\zenv=0.1$ planet of same mass. The effect of such an internal composition translates into a $\sim 8\%-9\%$ difference in the radius for this planet mass, whereas for Jupiter-mass or larger planets, the effect is $\lesssim 5\%$. Inferring the presence of significant heavy-element enrichment in the interior of exoplanets should
thus be in reach for transit planets smaller than about a Saturn-mass. Figure \ref{figpro} displays the interior pressure and density profiles of a $m_{\rm i}$=0.52 $\mjup$ planet, with or without core and/or heavy element enriched envelope.

Since the evaporation rates is inversely proportional to the mean planet density, ${\dot m}\propto (G\rho_P)^{-1}$ (see Baraffe et al. 2004, Eq. (1)), they are  affected by the presence of a core and an increase of $\zenv$. For the sequence with $m_{\rm i}$=0.52 $\mjup$ and with the structure predicted by the formation model ($\mcore$=6 $\mearth$ and $Z_{\rm env}$=0.1, solid line in Fig. \ref{fig2}), the sequence reaches masses $<$ 20 $\mearth$ (0.06 $\mjup$) within $\sim$ 6 Gyr with an escape rate $ \simeq 2 \times 10^{12}$ g/s ( $10^{-8}$ $\mearth$/yr) at this age
(see table \ref{summary}). Without a core and heavy element enrichment, this mass domain is reached within $\sim$ 1 Gyr. Fig. \ref{fig2} also reveals that when adopting the maximal evaporation rate, planets with initial masses $\sim$ 100 $\mearth$ ($0.3 \, \mjup$) reach the runaway phase described in Baraffe et al. (2004) and evaporate entirely within $\sim$ 100 Myr regardless of the presence or not of a core and heavy element enrichment. We must also recall that the evaporation rates depend on the high energy flux from the star, which decreases exponentially with time. This decrease, combined with the rather low average density of planets at early ages, yields escape rates at the beginning of the evolution which are $\sim$ 30 times higher than those obtained after 1 Gyr (see Baraffe et al. 2004).

\begin{table}
\caption{Effect of heavy element content on the radius of a 0.3 $\mjup$ ($\sim$ 100 $\mearth$) at different ages. Evolution proceeds at constant mass,
i.e $\dot M$ = 0. Case A corresponds to $\mcore=0$, $\zenv$=0.02; Case B
corresponds to $\mcore=6 \mearth$, $\zenv$=0.1. Irradiation effects are included
in both cases.}
\begin{tabular}{lcc}
\hline\noalign{\smallskip}
 age & R/$\rjup$ &  R/$\rjup$  \\
  (Gyr)   & (case A) & (case B)\\
\noalign{\smallskip}
\hline
\noalign{\smallskip}
0.5 &  1.19 & 1.09 \\
1 & 1.14 & 1.05 \\
5 & 1.06 & 0.98 \\
\hline
  \end{tabular}
   \label{struc}
  \end{table}

\subsection{Evolution with lower evaporation rates}

\begin{figure}
\psfig{file=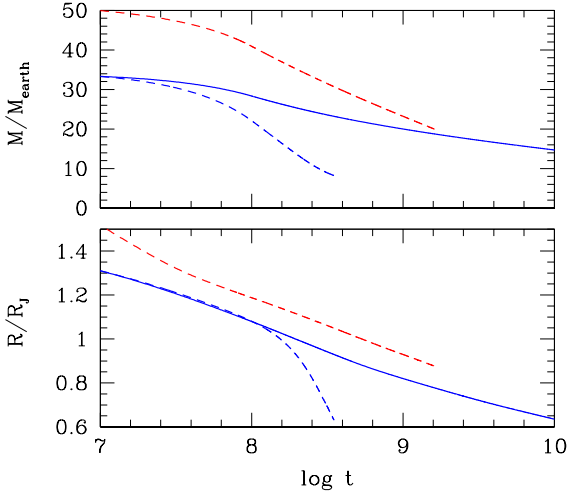,height=88mm,width=88mm}
\caption{Same as Fig. \ref{fig2} for planets with initial mass   50 $\mearth$ (0.15 $\mjup$)  and 33 $\mearth$ (0.1 $\mjup$) and escape rates 1/10 (dash) and 1/20 (solid) times the rates of L03.
For both cases, $\mcore$=6 $\mearth$.  For the 50 $\mearth$ planet (upper dashed curve in both panels), $Z_{\rm env}$=0.2 and for the  33 $\mearth$ planet, $Z_{\rm env}$=0.4 (cf. Fig. 1).}
\label{figL10}
\end{figure}

We now consider the effect of different evaporation rates, below the L03 value, on the evolution of the planets. Since
the present formation model (\S 2) can produce planets with masses smaller than  $\sim$ 100 $\mearth$ (0.3 $\mjup$), we consider various objects below this limit. 
Figure \ref{figL10} portrays the evolution of planets with initial masses of 50 $\mearth$ ($\sim$ 0.15 $\mjup$) and
33 $\mearth$ ($\sim$ 0.1 $\mjup$), respectively, for evaporation rates corresponding to 1/10 and 1/20 times
the L03 value. For the corresponding planet densities, these rates correspond to $\dot M \, \simle 10^{12}$ g/s ($6.8 \cdot 10^{-9}$ $\mearth$/yr) and $ \simle 5 \times 10^{11}$ g/s (2.7$\cdot 10^{-9}$ $\mearth$/yr) after 1 Gyr, respectively (see table \ref{summary}). As mentioned previously, such 10 to 20 smaller rates than the L03 maximum escape value are suggested by various calculations (Lecavelier des Etangs et al.  2004; Yelle 2004) and
in particular by the recent multi-layer hydrodynamical calculations of Tian et al. (2005). The planets have a 6 $\mearth$ core and a heavy element enrichment in the envelope of $Z_{\rm env}$=20\%  and $Z_{\rm env}$=40\% in mass fraction 
for initial masses 50 $\mearth$ and 33 $\mearth$ respectively (Fig. 1). As seen in Fig. \ref{figL10}, for such rates, the planets reach a Neptune-mass within a few hundred of million years to a few gigayears, meaning that about 2/3 to 1/2 of their initial mass has been lost under the influence of the parent star high energy flux.

\begin{figure}
\psfig{file=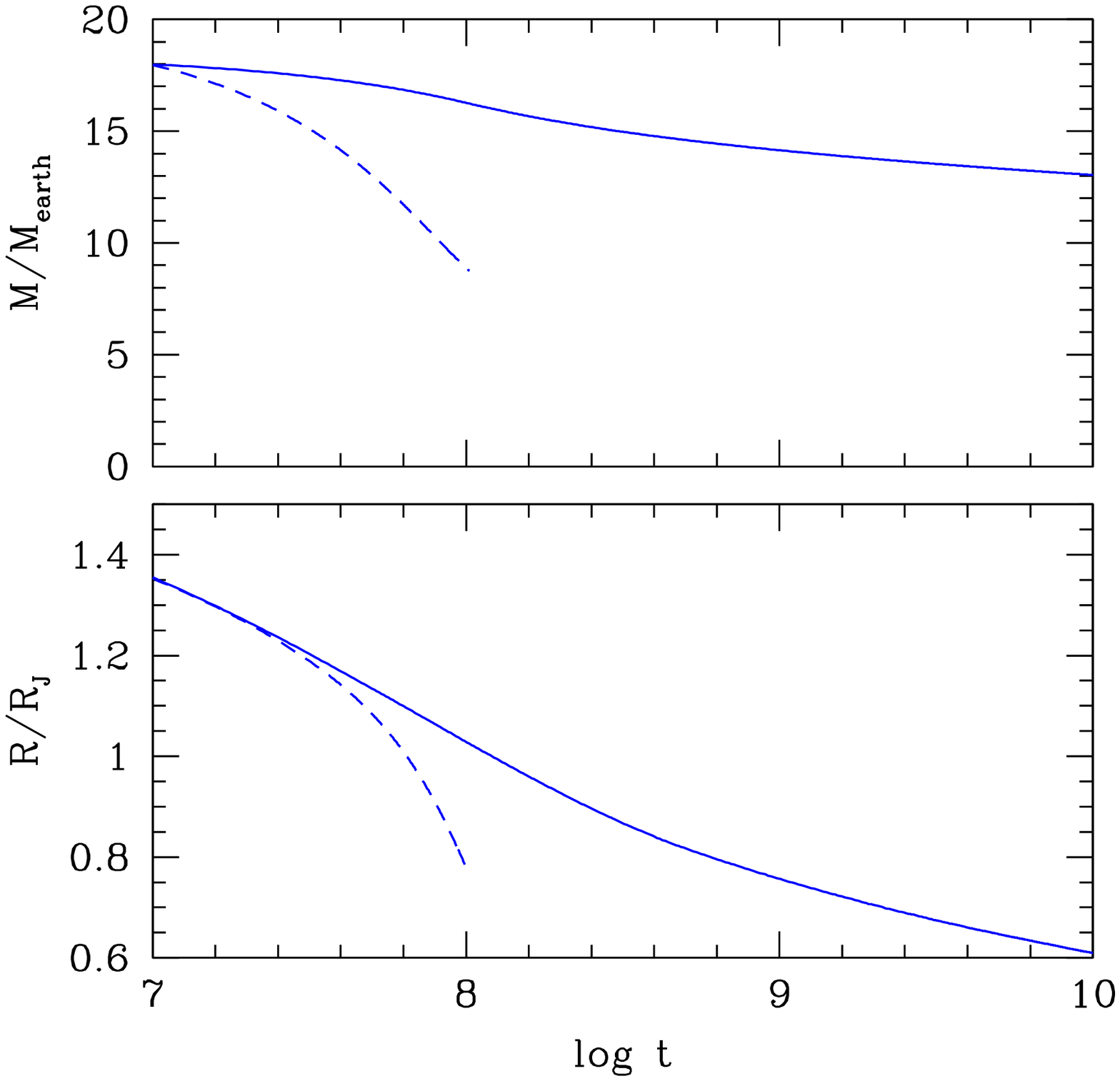,height=88mm,width=88mm} 
\caption{Same as Fig. \ref{fig2} for planets with initial masses  18 $\mearth$ (0.056 $\mjup$),
$\mcore$=6 $\mearth$ and   $Z_{\rm env}$=0.4 (see Fig. 1) but  different escape rates:  1/20 (dash) and 1/100 (solid) times the rates of L03.
}
\label{figL100}
\end{figure}

If we adopt an escape rate corresponding to 1\% of the L03 value, i.e. a reduction by a factor 100,
in order for the planet to reach a Neptune-mass within $\sim 1$ Gyr, the initial mass must be $m_{\rm i} \simeq 18 \mearth$ (0.056 $\mjup$), as illustrated in Fig. \ref{figL100}. This corresponds to a $\sim 20\%$ mass loss of the planet envelope.
For this initial mass, our formation models predict an envelope enrichment $Z_{\rm env} \simeq 0.4$.

This abundance of heavy elements strongly depends on the treatment of planetesimal accretion (size of the planetesimals, mechanical destruction in the envelope, convection in the envelope, see \S 5.2). To test how sensitive our evolutionary models are to these issues, we performed a comparison calculation in which we assumed that all the accreted planetesimals sink through the envelope without mass loss and fall onto the core. In such a model, a planet with an initial mass of 18 $\mearth$ is found to have a core mass of  $\sim$ 11 $\mearth$ surrounded by an envelope essentially composed of H and He. In that case, the evolutionary properties are very similar to the ones obtained with our nominal model (characterized by  $\mcore$=6 $\mearth$ and $Z_{\rm enve}$=0.4) for L03 evaporation rates divided by 100.  The radius found in the former case is $\sim$ 0.55 $\rjup$ at 10 Gyr, instead of $0.6$ $\rjup$ for the nominal model (solid curve in Fig. \ref{figL100}). The different evolutionary sequences are summarized in Table \ref{summary}.

\section{Predictions and uncertainties}

\subsection{Predictions}

Our results predict that if the formation and evolution of a hot-Neptune, such as the planet around $\mu$-Ara, proceeds as described by one of the sequences displayed in Figs. \ref{fig2}-\ref{figL100}, its radius should be greater than $\sim$ 0.6 $\rjup$, regardless of the escape rate adopted. Baraffe et al. (2005) reached similar conclusions for coreless evolutionary sequences and maximal escape rate. The present work confirms this prediction with more realistic planetary structures, consistent with the core accretion model scenario. 
It is important to understand the reason for such a large radius and thus to disentangle the various effects, evaporation, irradiation and heavy element enrichment. This is illustrated in Table \ref{neptune}
which displays a few values of the radius of Neptune mass planets evolved at constant mass with different $\zenv$ and with or without irradiation effects included. We first recall that evaporation effects do not affect the evolution, and thus the mass-radius
relationship of the planet as long as this latter is not in the evaporation runaway phase (see Baraffe et al. 2004). The values of radii given in Table \ref{neptune} are thus also characteristic of radii obtained for evolutionary sequences including evaporation for same mass and age (if they are not
in the runaway phase). As seen in the table (case 1 vs case 2), an increase of heavy element from $\zenv$=0.1 to $\zenv$=0.4
yields a $\sim$ 15\% decrease of the radius of an irradiated Neptune mass planet at 5 Gyr while,
for a given metal enrichement $\zenv$=0.4, irradiation effects increase
the radius by $\sim$ 20-30\% at a given age.
Table \ref{neptune} also shows that radii as large as 0.8 $\rjup$ after a few Gyr can be reached for an envelope metal fraction $\zenv$=0.1 in the irradiated case. This envelope metallicity is predicted by our formation model if the initial planet mass is $>$ 150 $\mearth$ ($\sim$ 0.6 $\mjup$). In order for such
a high progenitor mass to reach a Neptune mass 
planet within a few gigayears, high escape rates are required (see \S 4.1 and Fig. \ref{fig2}).
This analysis shows that the aforementioned $\sim$ 0.6 $\rjup$ lower limit for the radius of hot Neptunes stems essentially from the effect of irradiation on a planet which retains a substantial gaseous (H, He) envelope.

In contrast, if $\mu$-Ara-like planets are hot cores with small gaseous envelope, as suggested by Brunini \&  Cionco (2005), their radius should be significantly smaller, closer to the Neptune planet radius ($\sim$ 0.35 $\rjup$). Radius determinations of hot-Neptunes could thus distinguish between different formation scenarios. They may also provide information about the efficiency of the evaporation process, since measurements of radii larger than $\sim$ 0.8 $\rjup$, for planets older than a few Gyr, would indicate high initial
progenitor mass and thus high escape rates, as above mentioned.
Evaporation rates also affect the time spent in the mass range 10-20 $\mearth$ (0.03-0.06 $\mjup$). Evolution of planets at the maximal evaporation rate proceeds rapidly and the corresponding detection probability while the planet lies within this mass range is much smaller than for significantly reduced rates.

\begin{table}
\caption{Effect of heavy element content and of irradiation on the radius of a Neptune mass planet ($\sim$ 17 $\mearth$) at different ages. Evolution proceeds at constant mass,
i.e $\dot M$ = 0. All models have $\mcore=6 \mearth$. Case 1 corresponds to $\zenv$=0.1, with irradiation effects (from a Sun at 0.09 AU); Case 2 is similar to case 1 with  $\zenv$=0.4. Case 3 is with $\zenv$=0.4,
without irradiation effects.}
\begin{tabular}{lccc}
\hline\noalign{\smallskip}
 age & R/$\rjup$ &  R/$\rjup$  &  R/$\rjup$ \\
  (Gyr)   & (case 1) & (case 2) & (case 3) \\
\noalign{\smallskip}
\hline
\noalign{\smallskip}
1 &  0.93 &  0.80 & 0.62\\
5 &  0.83 & 0.70 & 0.58 \\
\hline
  \end{tabular}
   \label{neptune}
  \end{table}

The internal composition also strongly affects the evolution of the planet. The sequence with initial mass 0.52 $\mjup$ (166 $\mearth$), with $\mcore$=6 $\mearth$ and $Z_{\rm env}$=0.1 (solid line in Fig. \ref{fig2}), spends 1.1 Gyr with a mass $<$ 30 $\mearth$ (0.094 $\mjup$). For the same object without heavy element enrichment in the envelope (dashed line in Fig. \ref{fig2}), this time is reduced by a factor $\sim$ 4. Finally, for the same planet without a core and heavy element enrichment (dotted line in Fig. \ref{fig2}), the time is reduced by a factor $\sim$ 15. A planet with initial mass  1 $\mjup$,  with the same $\mcore$=6 $\mearth$ and  $Z_{\rm env}$=0.1, looses only 15\% of its total mass after 10 Gyr, using the rates of L03 at an orbital distance $a$ = 0.09 AU. Jupiter-like planets thus remain almost unaffected by evaporation at such orbital distances. Assuming the same formation probability for a 0.5 $\mjup$  and a 1 $\mjup$ planet, our model predicts {\it for the high L03 evaporation rate} a significantly lower detection probability for hot-Neptunes than for hot-Jupiters. Roughly 10 times smaller, comparing  the 1.1 Gyr lifetime of a planet with  mass $<$ 30 $\mearth$ to the 10 Gyr lifetime of  a Jupiter-mass planet.  Note that without heavy element enrichment (in the core and in the envelope), this probability would even be significantly smaller.

\subsection{Uncertainties and model improvements}

In the following, we describe a number of assumptions used in the core accretion model and the evolutionary calculations, which need to be improved in the future.

A main uncertainty in the formation model stems from the treatment of the accreted matter and its contribution to the gravitational potential of the planet. As done in Pollack et al. (1996),  the mass of accreted planetesimals is systematically added to the core mass in the gravitational potential calculation, whether they are destroyed in the envelope or fall onto the core (this is the so called "sinking approximation", see Pollack et al. 1996,  Alibert et al. 2005a). Work is underway to include a more consistent treatment of the gravitational potential, taking into account the distribution of the accreted matter within the envelope. Note, however, that planetesimals are destroyed very deep in the envelope\footnote{deeper than in former 
calculations by Pollack et al. 1996, see Alibert et al. 2005a}, in regions close to the core/envelope interface.

Concerning  evolutionary models, we have already discussed the assumption used for the treatment of  heavy element enrichment in the envelope (see \S 3.3). A more consistent treatment of such enrichment may affect the envelope structure and, hence, the mean
density of the planet and the evaporation rates. It is difficult to estimate the differences between an envelope structure based on the assumption of an effective helium fraction, as used here, and the one based on a consistent EOS. For values up to $Z\approx 10\%$ of heavy element enrichment in the envelope, however, we anticipate that a proper treatment of the equation of state should not affect drastically the present results.

Another important assumption stems from the irradiated atmosphere models used as outer boundary condition for the interior structure. These models (Barman et al. 2001, 2005) include
irradiation effects with up-to-date molecular and grain opacities, but have been computed for solar composition. Taking into account the increase of metallicity in atmospheric calculations up to levels inferred by the present formation models is a complex task and a long term project (see Chabrier et al. 2006 for preliminary results).
We thus use here solar composition atmospheric models. Note  however that for close-in planets, irradiation effects from the parent star yield planetary atmospheric structures which are radiative even at deep levels (Barman et al. 2001; Baraffe et al. 2003). Therefore, when the proto-planet accretes, the bulk of planetesimals either settle down to the core or are destroyed in the deep envelope. Only the smallest planetesimals may pollute the atmosphere. If convection in the interior is efficient, heavy elements will be mixed homogeneously, but should not pollute the outermost irradiated, radiative surface layers. This is no longer true, of course, for non-irradiated planets, as measurements of the Jupiter atmosphere by the Galileo probe indicate 
abundances of CH$_4$, H$_2$S, NH$_3$ three times solar. For hot-Neptunes and hot-Jupiters, we may thus expect the heavy element enrichment in the atmosphere to be less dramatic than
in the convective envelope. This point is also important for the treatment of evaporation, since the model of L03 applies to hydrogen dominated upper atmosphere. The effect of heavy element contamination on evaporation rates still needs to be explored. Intuitively, one expects that an enrichement in heavy elements yields a less efficient evaporation process. However, the effect is certainly more complex, since a change of the  chemical composition will affect the whole chemistry and cooling properties of the upper atmosphere.  The effects of heavy element enrichment on atmosphere structures and evaporation process are thus important problems to study in the future.

\section{Conclusion}

The present study presents first consistent calculations between planet formation, within an improved version of the core accretion model, and subsequent evolution, including irradiation and evaporation. These calculations can explain, among others, the existence of hot-Neptune planets, like the one recently discovered around $\mu$ Ara. The formation model constructed by Alibert et al. (2005a) allows the existence of planets with initial mass ranging from $\sim$ 14 $\mearth$ (0.044 $\mjup$) to $\sim$ 400 $\mearth$ (1.25 $\mjup$). The planets end up having a dense core of $\sim$ 6 $\mearth$, independently of the total mass, and heavy element enrichment in the envelope varying from 10\% to 80\% from the largest to the smallest mass. With maximal evaporation rates,  as predicted by the models of L03, the progenitor of gigayear old $\mu$ Ara-like  planets has an initial mass $\sim$ 0.5 $\mjup$. 
For evaporation rates a factor 10 to 20 smaller, the progenitor has a mass in the range $\sim 0.1$-0.15 $\mjup$ ($\sim$30-50 $\mearth$), meaning the planet has lost 1/2 to 2/3 of its initial envelope. For planets already forming in the Neptune-mass range, $\lesssim 20\,\mearth$, the L03 rate must be reduced by a factor 1/100 for the planets to survive to
evaporation at 0.09 AU from their Sun. In this latter case, the evaporation rates vary from $\sim$ 10$^{11}$ g/s (6 $10^{-10}$ $\mearth$/yr), at an age of 1 Gyr, to $\sim$ 7 $10^{9}$ g/s (4 $10^{-11}$ $\mearth$/yr) at 10 Gyr. Our current knowledge of evaporation processes  in exoplanet atmospheres is still too embryonic to favor or exclude any of these values, although recent hydrodynamical simulations favor the intermediate value (Tian et al. 2005). The range of escape rates explored in the present calculations, however, should bracket the "real" solution.

Our calculations provide an {\it observable diagnostic} for such a core-accretion-irradiation scenario for the formation and evolution of hot-Neptunes, namely radii $\simgr$  0.6 $\rjup$ for Neptune-mass planets. Also, the scenario with maximal evaporation rate may provide a statistical signature in terms of the number of hot-Neptunes, which is expected to be much smaller than the number of hot-Jupiter-like planets. In the absence of detailed statistical analysis,
a crude comparison of evolutionary timescales suggests 10 times  less hot-Neptunes than hot-Jupiters at  $a \simeq 0.09$ AU, {\it for the L03 maximal escape rate}. The detection of a larger fraction of hot-Neptunes at such orbital separation would thus favor scenarios with lower evaporation rates.

A complete understanding of the formation and evolution of light, Neptune-mass planets is still far from reach. The present attempt to derive a {\it consistent picture} from the planet genesis to its present day conditions provides one possible alternative path for the birth and fate of these objects. It also points to
several issues to be addressed
for a better modelling of planet formation and interior and atmospheric structures.
Further improvement requires in particular (i) a better treatment of heavy element enrichment, both in the envelope and in the atmosphere and (ii) a better understanding of evaporation processes.

In parallel with such theoretical developments, observations of planets down to a few Earth-masses and located at orbital separations large enough ($a \simgr 1 $ AU) to remain unaffected by evaporation would provide strong constraints on the formation scenarios of light planets. This is another appealing challenge for planet hunters.

\begin{acknowledgements}  I.B and G. C thank warmly the International 
Space Science Institute and the Physikalisches Institut of the Bern University
for hospitality during elaboration  of this work. Support from the Swiss National Science Foundation is gratefully acknowledged. The calculations were performed using the computer facilities of the Centre de Calcul Recherche et Technologie (CCRT) of CEA.
\end{acknowledgements}

{}

\end{document}